%% file: main.tex
\documentclass[10pt,conference,english,twocolumn]{IEEEtran} % conference two column

\usepackage{babel}
\usepackage[babel]{microtype}
\usepackage[babel]{csquotes}

\usepackage[T1]{fontenc}
\usepackage[svgnames]{xcolor}
\usepackage{graphicx}
\usepackage{tikz}
\usepackage{pgfplots}
\usepackage{subfigure}

\usepackage{tabularx}
\usepackage{booktabs}

\usepackage{amsmath}
\usepackage{amsfonts}
\usepackage{amssymb}
\usepackage{amsthm}
\usepackage{bm}
\usepackage{siunitx}
\usepackage{mathtools}
\usepackage{dsfont}

\usepackage[math]{blindtext}
\usepackage[linesnumbered,ruled]{algorithm2e}
\usepackage{algpseudocode}

\newtheorem{theorem}{Theorem}

\usepackage{hyperref}
\usepackage[acronyms,nomain,xindy,nowarn]{glossaries}
\makeglossaries
\loadglsentries{acronyms.tex}
\setacronymstyle{long-short}
\glsdisablehyper

\input{setup-colors.tex}

\input{setup-plots.tex}
\input{setup-misc.tex}

\title{{Multi-Connectivity for Multicast Video Streaming in Cellular Networks (Extended Abstract)}}
\author{%
	\IEEEauthorblockN{%
		Sadaf ul Zuhra~\IEEEauthorrefmark{2}, Prasanna Chaporkar~\IEEEauthorrefmark{1}, Abhay Karandikar~\IEEEauthorrefmark{1},  and H. Vincent Poor~\IEEEauthorrefmark{2}}
	\IEEEauthorblockA{\IEEEauthorrefmark{2}Department of Electrical and Computer Engineering, Princeton University, USA}
	\IEEEauthorblockA{\IEEEauthorrefmark{1}Department of Electrical Engineering, Indian Institute of Technology Bombay, Mumbai, India}
	Email: sadaf.zuhra@princeton.edu, $\lbrace${chaporkar, karandi}$\rbrace$@ee.iitb.ac.in, poor@princeton.edu
}

\begin{document}
\maketitle

%\todo{Four sentences:}
%\color{blue}
%1) State the problem.
%2) Say why it is an interesting problem.
%3) Say what your solution achieves.
%4) Say what follows from your solution.
\begin{abstract}\noindent\boldmath
% Video streaming applications account for about three fourths of the global mobile data traffic and are responsible for using a major portion of the network resources. During large scale live streaming events (such as the Super Bowl) where a very large audience is requesting video content simultaneously, video streaming can lead to network congestion and poor quality of service. For such services, multicast transmission is an excellent solution that can prevent network congestion while also improving the quality of service of the users. However, providing seamless connectivity to cellular users in multicast video streaming remains an open problem. To address this issue, this paper explores the potential of using multi-connectivity (MC) in wireless multicast streaming. Our results reveal that MC significantly improves the performance of multicast services, especially for cell edge users who often suffer from poor channel conditions. We prove that optimal resource allocation in MC multicast streaming is an NP-hard problem. Therefore, we propose a greedy approximation algorithm for this problem with an approximation factor of $(1-1/e)$. We also prove that no other polynomial-time algorithm can provide a better approximation.
% %
In video streaming applications especially during live streaming events (such as the Super Bowl), video traffic can account for a significant portion of network traffic and can lead to severe network congestion. During such events, multicast transmission can be used to avoid network congestion since the same video content is being streamed to multiple users simultaneously. However, providing seamless connectivity to cellular users in multicast streaming remains an open problem. To address this issue, this paper explores the potential of using multi-connectivity (MC) in wireless multicast streaming. Our results reveal that MC significantly improves the performance of multicast services, especially for cell edge users who often suffer from poor channel conditions. We prove that optimal resource allocation in MC multicast streaming is an NP-hard problem. Therefore, we propose a greedy approximation algorithm for this problem with an approximation factor of $(1-1/e)$. We also prove that no other polynomial-time algorithm can provide a better approximation.

%
% Multi-connectivity has emerged as a key enabler for providing seamless connectivity in cellular mobile networks. However, its potential for improving the quality of multicast transmissions has remained unexplored. In this paper, we investigate the use of multi-connectivity for wireless multicast streaming. Multi-connectivity can significantly improve the performance of multicast services. It especially benefits the cell edge users who often suffer from poor channel conditions. In this work, we assess the impact of multi-connectivity on the performance of multicast streaming. We propose procedures for establishing multi-connectivity in a multicast system and address the associated resource allocation problem. We prove that the optimal resource allocation problem is NP-hard. We propose a greedy approximation algorithm for this problem and prove that no other polynomial-time algorithm can provide a better approximation. Since video streaming is the primary use case under consideration here, we use traces from actual videos to generate realistic video traffic patterns in our simulations. Our simulation results clearly establish that multi-connectivity results in considerable performance improvement in multicast streaming.
\end{abstract}
\glsresetall

\section{Introduction}\label{sec:introduction}
Multicast refers to one to many transmissions in which a base station can use the same spectral resources to transmit content to multiple users simultaneously. This is especially useful for live streaming of content such as live telecasts of sports events, movie premiers, political events, and news telecasts. Using multicast for such services saves considerable network resources and enables serving a large number of users within a limited bandwidth~\cite{self,tech_report}.

Multi-connectivity (MC) allows users to connect to and receive content from multiple base stations and over multiple Radio Access Technologies (RATs) simultaneously. 
%Multi-connectivity is expected to be a key enabler in the Fifth Generation (5G) of wireless communication networks~\cite{sylla2022multi}. The high data rate, ultra-reliable low latency, and high mobility requirements of 5G necessitate the reduction of radio link failures due to mobility. Multi-connectivity makes it possible to avoid such failures and ensures seamless connectivity for mobile users.
% By allowing users to receive content from multiple base stations simultaneously, it allows serving a larger number of users and improves the performance of cell edge users.
Allowing multi-connectivity in multicast streaming~\cite{patent} further increases the serving capacity of a cell, reduces the dependence of multicast transmissions on the weakest users in the system, and makes the multicast streaming operations more robust. 
%Contributions:
% In this paper, we explore the use of multi-connectivity in multicast transmissions. We define procedures for establishing multi-connectivity for users in a multicast system. Since a multi-connected system involves users receiving content from multiple base stations simultaneously, the corresponding resource allocation problem needs to consider a global view of the system to make optimal allocation decisions. We formulate the resource allocation problem for this system with the objective of maximizing the total number of users served. We prove that this resource allocation problem is NP-hard and propose a centralized greedy approximation algorithm for solving it. Since centralized resource allocation incurs additional control overheads, we also propose a distributed allocation policy. We evaluate the performance improvements provided by multi-connectivity in multicast through extensive simulations. 
%
\section{System Model and Problem Formulation}\label{sec:system-model}
%
% \begin{figure}[!htb]
% \centering
% \includegraphics[scale = 0.5]{figures/mbms_architecture}
% \caption{MBMS architecture}
% \label{fig:mbms_architecture}
% \end{figure}
%
Consider a system of $C$ cells, each with a base station located at the center. There are $M$ multicast users in the system that are all capable of multi-connectivity and can potentially be served by any number of base stations.
The base station of the cell in which a user located is called as its primary base station.
It is assumed that all the multicast users are requesting the same content that is being multicast by all the base stations. The multicast stream has a rate requirement of $R$ bits/second. 
Content is streamed at this rate $R$ whenever the multicast session is active.  The users in a cell that are subscribed to the same multicast session form a single multicast group and receive the streaming content over the same physical resource blocks (PRBs). Users can potentially receive the content from any number of neighboring base stations in addition to their primary base station. A multi-connected user, therefore, belongs to multiple multicast groups streaming the same content. We assume that the multicast stream in a cell is allocated one PRB in each sub-frame. The multicast data stream from the primary and secondary base stations of a user may or may not be scheduled on the same PRB.
 % Resource allocation to various multicast streams can either be done by each base station independently or by a central controller that manages the base stations.  

 The channel states of users vary across time and frequency. As a result, a user experiences different channels in different sub-frames and across different PRBs in a sub-frame. Depending on the channel state of a user, there is a certain maximum rate it can successfully decode in a PRB. Since the multicast content is transmitted at rate $R$, a user may or may not be successfully served by the base station to which it is connected. For instance, consider that base station $c$ is streaming the multicast content over PRB $j$ in sub-frame $t$. Let $r^c_{jk}[t]$ be the maximum decodable rate for user $k$ in PRB $j$ of cell $c$ in sub-frame $t$. If $R > r^c_{jk}[t]$, user $k$ will not be able to successfully receive the content from base station $c$. On the other hand, if $R \leq r^c_{jk}[t]$, user $k$ is successfully served by base station $c$.
  A multi-connected user is said to be successfully served in sub-frame $t$ if it can decode the content from any of the base stations to which it is connected. On the other hand, a user that is not multi-connected would be served only if it can decode the content from its primary base station. 
  % The resource allocation problem for this system is define as follows. 
%
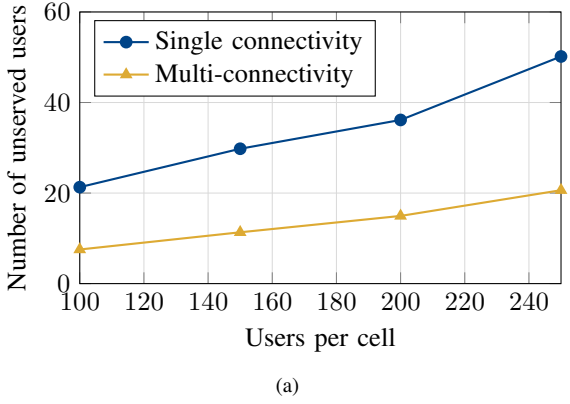
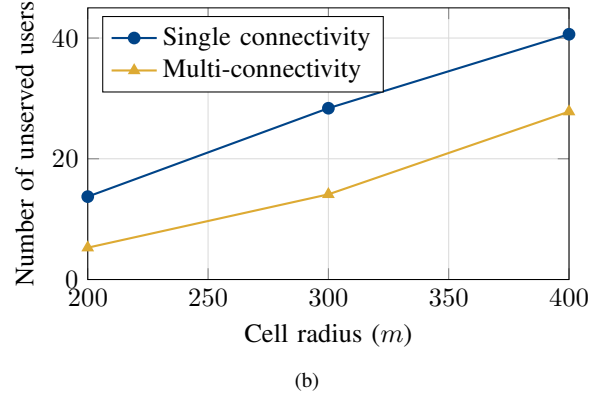
\begin{figure*}[t]
	\centering
	\subfigure[\label{Fig1a}]{%
		\input{img/group_size.tex}
	}
	\hfill
	\subfigure[\label{Fig1b}]{%
		\input{img/cell_size.tex}
	}
	\caption{Average number of unserved users under greedy approximation algorithm as a function of, a)\ number of users, b)\ cell radius.}
	\label{fig:results}
\end{figure*}

\subsection{Problem Formulation}\label{sub:problem-formulation}
The objective of the resource allocation problem in the MC multicast system under consideration is to maximize the number of users successfully served in each sub-frame. Since a multi-connected user can receive content from multiple base stations, its performance depends on the PRB allocation in multiple cells. Therefore, a resource allocation algorithm must optimize over all the cells involved in the multicast streaming. The optimal resource allocation problem is defined as follows.

Consider $M$ multicast users distributed across $C$ cells that can receive multicast content from all the base stations in their neighborhood. $[M] = \lbrace 1,2, \ldots, M \rbrace$ is the set of all users. There are $N$ PRBs available in each cell. For all $j \in \{1,2, \ldots, N\}$ and all $c \in \{1,2, \ldots, C\}$, denote by $U_{jc} \subseteq [M]$, the set of users that are successfully served if PRB $j$ is allocated to the multicast service in cell $c$. Set ${\cal U}_c = \{U_{1c},U_{2c}, \ldots, U_{Nc} \}$ is the sub-collection of such sets for cell $c$ and let ${\cal U} = \{ {\cal U}_1, \ldots , {\cal U}_C\}$. The resource allocation problem can then be stated as follows: 

${\bf K^\star}$ :
Given the set of all users $[M]$ and the collection of sets ${\cal U} = \{ {\cal U}_1, \ldots , {\cal U}_C\}$, determine ${\cal U}' \subseteq {\cal U}$ such that
% ${\cal U}' \in \arg\max_{|{\cal U}'| = C} |\bigcup_{U_{jc} \in {\cal C}'} U_{jc}|$
$|\bigcup_{U_{jc} \in {\cal U}'}U_{jc}|$ is maximized subject to $|{\cal U}'| = C$ and $|{\cal U}'\cap {\cal U}_c| =1, \ \forall \ c$.

% In the next section, we discuss the computational feasibility of this problem.
%
\section{Preliminary Results}
\subsection{${\bf K^\star}$ is NP-hard} \label{subsec:nphard_mc_multicast}
The optimal resource allocation problem ${\bf K^\star}$ is an NP-hard problem. We prove this by reduction from the maximum coverage problem (MCP)~\cite{mcp}. MCP is known to be an NP-hard problem and is defined as follows.
 MCP takes as input a universal set ${\cal S}$, a number $k$ and a collection of sets ${\cal T} = \{T_1,T_2, \ldots, T_m\}$ where, for all $j \in \{1,2, \ldots, m\}$, $T_j \subseteq {\cal S}$. The objective of MCP is to determine a sub-collection ${\cal T}' \subseteq {\cal T}$ such that
 ${\cal T}' \in \arg\max_{|{\cal T}'| \leq k} |\bigcup_{T_j \in {\cal T}'} T_j|$.

\begin{theorem} \label{theorem:np_hard}
 ${\bf K^\star}$ is an NP-hard problem.
\end{theorem}
The proof of Theorem~\ref{theorem:np_hard} will be presented in the full version of this paper.

Since ${\bf K^\star}$ is NP-hard, no polynomial time solution exists for determining the optimal solution of ${\bf K^\star}$. Therefore, we propose the following approximation algorithm for the resource allocation problem under consideration.
\subsection{Approximation Algorithm for ${\bf K^\star}$} \label{sec:greedy_mc_multicast}
We construct a centralized greedy approximation (CGA) algorithm for solving ${\bf K^\star}$. The pseudo-code for this algorithm is given in Algorithm~\ref{algo:greedy}. CGA works iteratively by maximizing the number of additional users served in each iteration. In the first iteration, CGA chooses the set $U_{jc} \in {\cal U}$ that serves the maximum number of users. In each subsequent step, a set $U_{j'c'}$ is chosen that serves the maximum number of the remaining unserved users. In each step, the set chosen is from a different sub-collection ${\cal U}_c$ i.e., $c$ in the subscript of the chosen sets is different in each iteration. The collection of $C$ sets ${\cal U}_G$ thus chosen after $C$ iterations, provides the resource allocation for the system.

In the following theorem, we prove that the resulting solution has an approximation factor of $\left(1-\frac{1}{e}\right)$. This means that, the solution provided by this approximation algorithm serves at least $\left(1-\frac{1}{e}\right)$ of the number of users that would be served by the optimal algorithm.  \par

\begin{theorem} \label{theorem:approximation}
 The CGA algorithm (Algorithm~\ref{algo:greedy}) is a $\left(1-\frac{1}{e}\right)$ approximation for ${\bf K^\star}$. Furthermore, no other algorithm can achieve a better approximation unless P = NP.
 \end{theorem}
The proof of Theorem~\ref{theorem:approximation} will be presented in the full version of this paper.
\begin{algorithm}
	\KwIn{ Universe $[M]$, ${\cal U} = \{ {\cal U}_1, \ldots , {\cal U}_C\}$, $C$ }
	Initialize: ${\cal U}_G = \phi$\\
	\For{$n = 1:C$}{
	Pick $U_{j^\star c^\star} \in {\cal U}$ that covers the maximum number of elements from $[M] \setminus \bigcup_{U_{jc} \in {\cal U}_G}U_{jc}$ \\
	${\cal U}_G \leftarrow {\cal U}' \bigcup \{U_{j^\star c^\star}\}$ \\
	${\cal U} \leftarrow {\cal U} \setminus {\cal U}_{c^\star}$
	}
	\caption{Greedy Approximation Algorithm for ${\bf K^\star}$}
	\label{algo:greedy}
\end{algorithm}

\subsection{Numerical Results}
In Figure~\ref{fig:results}, the performance of the proposed MC multicast mechanism is illustrated through the total number of users left unserved in a seven cell multicast system. We compare the performance of MC multicast with single connectivity multicast in which each user is served by only its primary base station. It is observed that the number of unserved users decreases significantly with the use of MC in multicast streaming. Furthermore, from Figure~\ref{Fig1b}, it is also observed that MC multicast performs much better as the radii of the cells are increased which demonstrates its efficacy in serving the cell edge users.

\clearpage
\bibliographystyle{IEEEtran}
\bibliography{myrefs}

\end{document}

%% file: acronyms.tex
% !TeX spellcheck = en_US
\newacronym{5g}{5G}{Fifth-Generation Wireless Systems}
%A
\newacronym{ai}{AI}{Artificial Intelligence}
\newacronym{ann}{ANN}{Artificial Neural Network}
\newacronym{ask}{ASK}{amplitude-shift keying}
\newacronym[plural={AWGN}, \glsshortpluralkey={AWGN}]{awgn}{AWGN}{additive white Gaussian noise}

%B
\newacronym{bec}{BEC}{binary erasure channel}
\newacronym{ber}{BER}{bit error rate}
\newacronym{bler}{BLER}{block error rate}
\newacronym{bpsk}{BPSK}{binary phase-shift keying}
\newacronym{bs}{BS}{base station}

%C
\newacronym{cdf}{CDF}{cumulative distribution function}
\newacronym{cnn}{CNN}{convolutional neural network}
\newacronym{csi}{CSI}{channel state information}
\newacronym{csit}{CSI\babelhyphen{nobreak}T}{channel-state information at the transmitter}
\newacronym{csir}{CSI\babelhyphen{nobreak}R}{channel-state information at the receiver}

%E
\newacronym{ecc}{ECC}{error-correcting code}
\newacronym{ecdf}{ECDF}{empirical distribution function}
\newacronym{ee}{EE}{energy efficiency}

%F
\newacronym{ff}{FF}{feed-forward}
\newacronym{ffnn}{FF-NN}{feed-forward neural network}
\newacronym{fsk}{FSK}{frequency-shift keying}

%G
\newacronym{gm}{GM}{Gaussian mixture}

%I
\newacronym{iid}{i.i.d.}{independent and identically distributed}
\newacronym{il}{IL}{information leakage}
\newacronym{iot}{IoT}{internet of things}

%L
\newacronym{lb}{LB}{lower bound}
\newacronym{los}{LoS}{line-of-sight}

%M
\newacronym{mac}{MAC}{multiple access channel}
\newacronym{map}{MAP}{maximum a posteriori}
\newacronym{mc}{MC}{Monte Carlo}
\newacronym{mimo}{MIMO}{multiple-input multiple-output}
\newacronym{miso}{MISO}{multiple-input single-output}
\newacronym{ml}{ML}{machine learning}
\newacronym{mmwave}{mmWave}{millimeter wave}
\newacronym{mop}{MOP}{multi-objective programming problem}
\newacronym{mrc}{MRC}{maximum ratio combining}
\newacronym{msc}{MSC}{modulation and coding scheme}
\newacronym{mse}{MSE}{mean squared error}

%N
\newacronym{nlos}{NLoS}{non-line-of-sight}
\newacronym{nn}{NN}{neural network}
\newacronym{nve}{NVE}{normalized validation error}

%P
\newacronym{pdf}{PDF}{probability density function}
\newacronym{pwc}{PWC}{polar wiretap code}

%Q
\newacronym{qam}{QAM}{quadrature amplitude modulation}

%R
\newacronym{ra}{RA}{rearrangement algorithm}
\newacronym{rbf}{RBF}{radial basis function}
\newacronym{relu}{ReLU}{rectified linear unit}
\newacronym{ris}{RIS}{reconfigurable intelligent surface}
\newacronym{rl}{RL}{reinforcement learning}
\newacronym{rnn}{RNN}{recurrent neural network}

%S
\newacronym{sac}{SAC}{soft actor-critic}
\newacronym{sgd}{SGD}{stochastic gradient descent}
\newacronym{sgp}{SGP}{secure goodput}
\newacronym{simo}{SIMO}{single-input multiple-output}
\newacronym{sinr}{SINR}{signal-to-interference-plus-noise ratio}
\newacronym{siso}{SISO}{single-input single-output}
\newacronym{snr}{SNR}{signal-to-noise ratio}
\newacronym{svm}{SVM}{support vector machine}

%T
\newacronym{tvd}{TVD}{total variation distance}

%U
\newacronym{uav}{UAV}{unmanned aerial vehicle}
\newacronym{ub}{UB}{upper bound}
\newacronym{urllc}{URLLC}{ultra-reliable low latency communication}

%V
\newacronym{v2v}{V2V}{vehicle-to-vehicle}
\newacronym{v2x}{V2X}{vehicle-to-everything}

%W

%X

%Y

%Z
\newacronym{zoc}{ZOC}{zero-outage capacity}

%% file: setup-colors.tex
\definecolor{plot0}{HTML}{004488}
\definecolor{plot1}{HTML}{DDAA33}
\definecolor{plot2}{HTML}{BB5566}
\definecolor{plot3}{HTML}{000000}
\definecolor{plot4}{HTML}{AAAAAA}

%% file: setup-plots.tex
\pgfplotsset{compat=newest}
\pgfplotscreateplotcyclelist{lineplot cycle}{ %
	{plot0, mark=*, thick, mark options=solid},
	{plot1, mark=triangle*, thick, mark options=solid},
	{plot2, mark=square*, thick, mark options=solid},
	{plot3, mark=diamond*, thick, mark options=solid},
	{plot4, mark=pentagon*, thick, mark options=solid},
}

\pgfplotsset{%
	betterplot/.style={
		width=.93\linewidth,
		height=.27\textheight,
		xlabel near ticks,
		ylabel near ticks,
		cycle list name=lineplot cycle,
		mark options=solid,
		xmajorgrids=true,
		xminorgrids=true,
		ymajorgrids=true,
		grid style={line width=.1pt, draw=gray!20},
		major grid style={line width=.25pt,draw=gray!30},
		legend cell align=left,
		legend style = {
			/tikz/every even column/.append style={column sep=0.33cm}
		},
	},
}

%% file: setup-misc.tex
\usetikzlibrary{positioning,calc,external}
\newcommand{\todo}[2][]{\ignorespaces
	\if\relax\detokenize{#1}\relax
	{\color{red}[TODO: #2]}%
	\else
	{\color{red}[TODO (#1): #2]}%
	\fi
}

\definecolor{change}{HTML}{0096b8}

\theoremstyle{plain}% default  

\theoremstyle{definition}
\newtheorem*{prob*}{Problem Statement}

\theoremstyle{remark}

\newtheorem*{rem*}{Remark}

\newtheoremstyle{example}{\topsep}{\topsep}{}{}{\itshape}{.}{ }{}
\theoremstyle{example}

\newtheorem*{example*}{Example}

\addto\extrasenglish{%

}

%% file: img/group_size.tex
\begin{tikzpicture}% function
	\begin{axis}[
		betterplot,
		width= .44 \linewidth,
		height=.22\textheight,
		xlabel={Users per cell},
		ylabel={Number of unserved users},
		legend pos=north west,
		xmin=100,
		xmax=250,
		ymin=0,
        ymax = 60,
		]
		\addplot+[mark repeat=1] 
        table[x=users,y=SC, col sep=comma] {data/unserved_groupsize.csv};
		\addlegendentry{Single connectivity};
		
		\addplot+[mark repeat=1] 
        table[x=users,y=MC, col sep=comma] {data/unserved_groupsize.csv};
		\addlegendentry{Multi-connectivity};
	\end{axis}
\end{tikzpicture}

%% file: img/cell_size.tex
\begin{tikzpicture}% function
	\begin{axis}[
		betterplot,
		width= .44 \linewidth,
		height=.22\textheight,
		xlabel={Cell radius ($m$)},
		ylabel={Number of unserved users},
		legend pos=north west,
		xmin=200,
		xmax=400,
		ymin=0,
        ymax = 45,
		]
		\addplot+[mark repeat=1] 
        table[x=radius,y=SC, col sep=comma] {data/unserved_cells.csv};
		\addlegendentry{Single connectivity};
		
		\addplot+[mark repeat=1] 
        table[x=radius,y=MC, col sep=comma] {data/unserved_cells.csv};
		\addlegendentry{Multi-connectivity};
	\end{axis}
\end{tikzpicture}